\documentclass{article}

\usepackage[english]{babel}

\usepackage[letterpaper,top=2cm,bottom=2cm,left=3cm,right=3cm,marginparwidth=1.75cm]{geometry}

\usepackage{amsmath}
\usepackage{amssymb}
\usepackage{amsfonts}
\usepackage{graphicx}
\usepackage[colorlinks=true, allcolors=blue]{hyperref}

\begin{document}

\begin{center}
\large{\bf New envelope equations for shallow water waves and
modulational instability}
\end{center}
\vspace{0.8 cm}
\begin{center}
{\bf A. Marin (1, 2), A. S. Carstea (2)}\footnote{email: acarst@theor1.theory.nipne.ro, 
carstea@gmail.com}\\
\vspace{0.4 cm}
{\it (1) Department of Physics, University of Bucharest,}\\
{\it Magurele, P.O. Box MG11 Bucharest, ROMANIA}\\
\vspace{0.1 cm}
{\it (2) Department of Theoretical Physics, Institute of Physics and Nuclear
Engineering,}\\
{\it Magurele, P.O. Box MG6 Bucharest, ROMANIA}\\
\end{center}
\vspace{1 cm}
\abstract{The dynamics of wave groups is studied for long waves, using the framework of the Benjamin-Bona-Mahony (BBM) equation and its generalizations.
It is shown that the dynamics are richer than the corresponding results obtained just from the Korteweg–de Vries-type equation.
First, a reduction to a nonlinear Schrödinger equation is obtained for weakly nonlinear wave packets, and it is demonstrated
that either the focusing or the defocusing case can be obtained. This is in contrast to the corresponding reduction for the
Korteweg–de Vries equation, where only the defocusing case is found. The focusing regime displays modulational instability responsible for the appearance of rogue waves. Next, the condition for modulational instability is
obtained in the case of one and two monochromatic waves in interaction at slow space-time coordinates with equal scalings. Other new envelope equations are obtained starting from the general system describing shallow water waves found by Bona et al.\cite{bona}. A presumably integrable system is obtained form the integrable Kaup-Boussinesq one.

\section{Introduction}
Modulation of wave trains is an extremely important phenomenon since, at large scales, the emerging envelope equations display a richer phenomenology. For example in the case of deep water assuming weak nonlinearity the dynamics of the wave train are given by the famous {\it focusing} nonlinear Schrödinger equation (fNLS) and the formation of various coherent structures is triggered by the modulational instability.
If a quasi-monochromatic wave is propagating through a dispersive and weakly nonlinear medium then an instability of its amplitude appears against weak modulations with wave numbers lower than some critical values. The long time evolution leads to growth of the side bands and the energy exchange between these side bands reduces
the dispersion effects. When the dispersion and nonlinearity balance each
other coherent structures (e.g.rogue waves, solitons, breathers) may appear provided the envelope of the original equation is slower than the carrier wave \cite{grim}. Rogue waves which are rational solutions of fNLS correspond in reality to large amplitude (should exceed the significant wave height more than two times) waves appearing on the sea surface under almost normal conditions. The appearance of such waves is accompanied by deep ``holes” in the sea located nearby the giant wave \cite{kharif}, \cite{os}, \cite{dys}.

However the situation is different in the case of shallow water waves. Here the enevelope equations are usually given by {\it defocusing} nonlinear Schrödinger equation (dNLS) which is stable against weak modulation. For two-interacting waves described by the coupled defocusing NLS (Manakov system) there may be a possibility of instability.\\
In this paper motivated by a result obtained in atmospheric science we are trying to find some envelope equations staring from shallow water models different from KdV (or Boussinesq). The motivation of this paper is due to the results of Yano and Plant \cite{yano} about Benjamin-Bona-Mahony (BBM) equation describing atmospheric circulation. We intend to find the corresponding envelope equation and see the possible modulational instability which in turn may give some hints on extreme events in atmospheric dynamics. In addition we will study the generalisations of shallow water wave equations given by Bona et al.\cite{bona} and find the envelope correspondent. As an interesting off-shoot we found a new (possible) integrable envelope system which is dispersionless and (coming from Kaup-Boussinsq completely integrable system).\\
The paper is organised as follows; in the first section we show how to obtain the fNLS and dNLS starting form BBM equation. Then in the next section, assuming equal scales for space and time, we find a truncated equation of Yajima-Oikawa type describing also modulated waves, one component and two-components in interaction. We used the procedure given in \cite{onorato}. We show here that modulational instability is present. In the next two sections we start from a general system describing shallow water waves proposed by Bona et al.\cite{bona} (which contains the cases KdV, BBM, Kaup-Boussinesq etc) and find a general two-component enevelope system. Using Hirota bilinear formalism (here the form will be multi-linear due to non-integrability) we compute the solitary wave solution.  

\section{From BBM equation to nonlinear Schrödinger equation}

The BBM equation under consideration is
\begin{equation}\label{bbm}
V_{\tau}+b V_{\xi}+VV_{\xi}-aV_{\xi\xi\tau}=0
\end{equation}
It comes from the model of Plant-Yano on atmospheric dynamics with a non-zero constant large scale circulation \cite{yano}
The BBM dispersion relation is
$$\omega=-\frac{bk}{1+ak^2}$$

By scaling $\tau$, $V$ and $a$ we can put $b=1$. Considering a quasimonochromatic wave as the effect of small nonlinearity we put
$$V(\xi,\tau)=\sum_{n\in{\mathbb Z}}\epsilon^{\alpha_n}v_n(X,T)e^{in(k\xi+\omega\tau)}$$
with the following stretched variables
$$X=\epsilon(\xi-w\tau), T=\epsilon^2\tau$$
$w$ being a velocity to be determined.
At order $O(\epsilon^3)$ and $n=0$ we find
$$v_0=\frac{|v_1|^2}{w-b}$$
At order $n=2$ and $O(\epsilon^2)$ we find
$$v_2=\frac{1+ak^2}{6abk^2}v_1^2$$
At $n=1$ the order $O(\epsilon)$ is identically satisfied. The next order $O(\epsilon^2)$ is an equation which gives the speed $w$
$$w=b\frac{1-ak^2}{(1+ak^2)^2}$$
At order $O(\epsilon^3)$ we find the nonlinear Schrödinger equation
\begin{equation}
 i\partial_Tv_1+\frac{ab(5ak^2+3)}{6abk(ak^2+3)}|v_1|^2v_1+\frac{kab(ak^2-3)}{(1+ak^2)^3}\partial_X^2v_1=0
\end{equation}

{\bf Remarks:}
\begin{itemize}
 \item If $ak^2-3<0$ the nonlinear Schrödinger equation is defocusing. In this case $k<\sqrt{3}/a$ so we are in a regime with long waves (limit where BBM behaves as KdV)
 \item If $k>\sqrt{3}/a$ then we are in the focusing regime of nonlinear Schrödinger which displays modulational instability.
\end{itemize}
If we make the following notations
$$A=\frac{kab(ak^2-3)}{(1+ak^2)^3}, \quad B=\frac{3+5ak^2}{6abk(3+ak^2)}$$
and the scalings $v_1\to v_1/\sqrt{B}, X\to X/\sqrt{2A}$ we can write the NLS equation in a canonical form:
$$i\partial_Tv_1+|v_1|^2v_1+\frac{1}{2}\partial_X^2v_1=0$$
The solutions can be found using the bilinear formalism. Considering 
$v_1=G/F$ ($G$-complex, $F$-real) and introducing the Hirota bilinear operator
$$D_x^n a\cdot b=(\partial_y)^na(x-y)b(x+y)|_{y=0}$$
we have the following bilinear form
$$(iD_t+(1/2)D_x^2)G\cdot F=0$$
$$D_x^2F\cdot F-|G|^2=0$$
having the rogue-wave solution (Peregrine breather)
$$v_1(X,T)=e^{iT}\left(1-\frac{4(1+2iT)}{1+4X^2+4T^2} \right)$$
and the final rogue solution solution of the BBM (up to order $\epsilon^3$) 
$$V(\xi,\tau)=\epsilon v_1(X,T)e^{i\theta}+\epsilon^2\left(\frac{|v_1(X,T)|^2}{w-b}+\frac{1+ak^2}{6abk^2}v_1(X,T)^2 e^{2i\theta}\right)+O(\epsilon^3)$$
where $\theta=k\xi+\omega\tau, w=-bk/(1+ak^2).$

The NLS equation has also the following breather (time-periodic) type solutions:
$$v_1(X,T)=e^{2iT}\frac{\cos(\Omega T-2i\phi)-\cosh{\phi}\cosh{(pX)}}
{\cos(\Omega T)-\cosh(\phi) \cosh(pX)}
$$
where $p=2\sinh\phi, \Omega=2\sinh(2\phi)$, $\phi$ is a free real parameter. Also we have the space-periodic breather
$$
v_1(X,T)=e^{2iT}\frac{\cosh(\Omega T-2i\phi)-\cos{\phi}\cos{(pX)}}
{\cosh(\Omega T)-\cos(\phi) \cos(pX)}
$$
\section{Interaction of two quasimonochromatic waves in BBM}
One can wonder what happens with the nonlinear interaction of waves propagating in the same direction in shallow water
characterized by a double-peaked power spectrum. 
The starting point is again the BBM equation and consider interaction of two quasimonochromatic waves peaked at $k_1$ and $k_2$ ($k_1\neq k_2$) in the slow variables $X=\epsilon x, T=\epsilon t$ (we follow the same procedure described in \cite{onorato} for KdV equation)
$$V(x,t,X,T)=\frac{\epsilon}{2}(Ae^{i\theta_1}+Be^{i\theta_2})+\frac{\epsilon^2}{2}(A_2 e^{2i\theta_1}+B_2e^{2i\theta_2}+C_2e^{i(\theta_1+\theta_2)}+D_2e^{i(\theta_1-\theta_2)}+\Phi)+c.c.$$
where $\theta_j=k_jx+\omega_jt, j=1,2$. Collecting terms proportional to the modes we obtain the following general {\it truncated}(not rigorous) system (we take $\epsilon=1$):
$$\mu_1 A_T+\mu_2A_X-ik_1s_0 A|B|^2+\frac{ik_1}{2}A\Phi-ik_1aA_{XT}+\mu_3A_{XX}=0$$
$$\nu_1 B_T+\nu_2B_X-ik_2s_0B|A|^2+\frac{ik_2}{2}B\Phi-ik_2aB_{XT}+\nu_3B_{XX}=0$$
$$\Phi_T+b\Phi_X+\frac{1}{4}(|A|_X^2+|B|_X^2)=0$$
where
$$
\mu_1=\frac{1+ak_1^2}{2},\quad \mu_2=\frac{b(1-ak_1^2)}{2(1+ak_1^2)},\quad \mu_3=\frac{ik_1ab}{2(1+ak_1^2)}
$$
$$
\nu_1=\frac{1+ak_2^2}{2},\quad \nu_2=\frac{b(1-ak_2^2)}{2(1+ak_2^2)}, \quad \nu_3=\frac{ik_2ab}{2(1+ak_2^2)}
$$
$$
s_0=\frac{(1+ak_1^2)(1+ak_2^2)}{8b((ak_1^2+3)^2+a(ak_1^2+6)k_2^2+a^2k_2^4)}
$$
The system looks like a generalisation of a bicomponent long-wave:short-wave resonance (sometimes called Yajima-Oikawa). This system displays modulational instability. To see this it is sufficient to consider one-component system (since the bi-component has larger freedom) namely,
\begin{equation}\label{yo1}
\mu_1 A_T+\mu_2A_X-ik_1s_0 A|A|^2+\frac{ik_1}{2}A\Phi-ik_1aA_{XT}+\mu_3A_{XX}=0\\
\end{equation}
\begin{equation}\label{yo2}
\Phi_T+b\Phi_X+\frac{1}{2}(|A|_X^2)=0
\end{equation}
This system is an extension of the one studied by Wadati, Segur and Ablowitz \cite{WSA}. To see the types of solitons we consider $A=G/F, \Phi=\partial_X^2\log F$ then the Hirota bilinear form of the system is given by:
$$P_1(D_T, D_X)G\cdot F=0,\qquad P_2(D_T, D_X)F\cdot F=P_3(D_T,D_X)G^*\cdot G$$
where Hirota polynomials are:
$$P_1(D_t,D_X)\equiv \mu_1 D_T+\mu_2 D_X-ik_1aD_{XT}^2+\mu_3D_X^2$$
$$P_2(D_t,D_X)\equiv D_{XT}^2+bD_{X}^2$$
$$P_3(D_T,D_X)\equiv 1$$

Two types of solitons appear here.

The first one:
$$F=1+Ke^{\eta+\eta^*}, \quad G=e^{\eta}, \quad \eta=pX+\omega T+\eta_0$$
where the dispersion relation and the phase $K$ are given by
$$P_1(p,\omega)=0,\quad K=\frac{P_3(p-p^*,\omega-\omega^*)}{2P_2(p+p^*,\omega+\omega^*)}$$

The second one (is trivial but in interaction with the first, in the integrable case, is nontrivial)
$$F=1+e^{\eta}, \quad G=0, \quad P_2(p, \omega)=0$$

The existence of two-soliton solution is valid if \cite{hie}:
$$P_2(p_1-p_2,\omega_1-\omega_2)P_1(p_1+p_2+p_1^*,\omega_1+\omega_2+\omega_1^*)+P_2(p_2+p_1^*,\omega_2-\omega_1^*)P_1(p_2-p_1^*+p_1,\omega_2-\omega_1^*+\omega_1)$$$$+P_2(p_1-p_1^*,\omega_1-\omega_1^*)P_1(p_1-p_2-p_1^*,\omega_1-\omega_2-\omega_1^*)=0$$
Unfortunately this condition is not satisfied by our system and accordingly the system does not have two soliton solution (i.e.non-integrability). 

\subsection{Modulational Instability}
Here we are going to discuss modulational instability \cite{li,bf} for the system (\ref{yo1}), (\ref{yo2}). The simplest monochromatic nonlinear wave is ($a_0$ and $m_0$ are {\it real} constants)
$$A=a_0 e^{i\Omega T}, \Phi=m_0, \quad \Omega=\frac{k_1s_0}{\mu_1}a_0^2-\frac{k_1m_0}{2\mu_1}$$
The nonlinear character can be seen from the fact that frequency depends on amplitude (Stokes wave).
We are interested in the stability of this simple nonlinear wave. Let us perturb it
$$A=(a_0+\zeta(X,T))e^{i(\frac{k_1s_0}{\mu_1}a_0^2-\frac{k_1m_0}{2\mu_1})T}$$
$$\Phi=m_0+z(X,T)$$
where $\zeta(X,T)$ is complex and $z(X,T)$ is real; both of them are considered small. Linearizing around them we get a linear system. Splitting the system in real and imaginary parts (by considering $\zeta=\zeta_R+i\zeta_I$) we obtain three coupled linear equations. Assuming 
$$\zeta_R=c_1 e^{i(PX-\sigma T)}, \zeta_I=c_2 e^{i(PX-\sigma T)}, z(X,T)=c_3 e^{i(PX-\sigma T)}$$ 
the compatibility condition gives the dispersion relation $\sigma=\sigma(P)$ as a solution of a cubic equation with complicated coefficients.
However for $k_1=a_0=b=1$ the discriminant has the form
$-(P^2-1)^2P^6 h(P)=0$ where h(P) is a polynomial of degree 6. For $P>3.2$, $h(P)$ is positive and accordingly the discriminant is negative, so $\sigma$ has {\it imaginary} part. This means that the monochromatic wave is unstable and there is posibility of rogue waves.

\section{General envelope system for shallow water waves}

We start from the following system of equations introduced in \cite{bona} to describe shallow water waves. .
\begin{equation}
    \begin{split}
       &\eta_t+u_x+(u\eta)_x+au_{xxx}-b\eta_{xxt}=0\\
       &u_t+\eta_x+uu_x+c\eta_{xxx}-du_{xxt}=0
    \end{split}
    \label{sNLS:bona}
\end{equation}
For various values of $a,b,c,d$ one can have Boussinesq system, Kaup-Boussinesq system, coupled KdV, coupled BBM etc. We consider solutions of the form $u(x,t)=A\exp{i(kx-\omega t)}$ and $\eta(x,t)=B\exp{i(kx-\omega t)}$. The dispersive part of the system rewrites as:
\begin{equation}
    \begin{split}
        & -i e^{i (k x-t \omega )} \left(B k \left(a k^2-1\right)+A \omega  \left(b k^2+1\right)\right)=0\\
        &-i e^{i (k x-t \omega )} \left(A k \left(c k^2-1\right)+B \omega  \left(d k^2+1\right)\right)=0
    \end{split}
\end{equation}
We solve this system for $\omega$ to obtain the dispersion relation:
\begin{equation}
    \omega^2=\frac{k^2 \left(1- a k^2\right) \left(1- c k^2\right)}{\left(b k^2+1\right) \left(d k^2+1\right)}
\end{equation}
By performing a series expansion around $k=0$ we obtain the large wavelength (small $k$) limit:
\begin{equation}
    \omega=k+\frac{1}{2} k^3 (-a-b-c-d)+\frac{1}{8} k^5 \left(-a^2+2 a b+2 a c+2 a d+3 b^2+2 b c+2 b d-c^2+2 c d+3
   d^2\right)+\mathcal{O}\left(k^6\right)
\end{equation}
Now we consider solutions for eq. (\ref{sNLS:bona}) of the form:
\begin{equation}
    \begin{split}
        &u(x,t)=\sum\limits_{n\in\mathbb{Z}}u_n e^{in\theta}\\
        &\eta(x,t)=\sum\limits_{n\in\mathbb{Z}}\eta_n e^{in\theta}
    \end{split}
\end{equation}
Here $\theta=kx-kt+\frac{1}{2}(a+b+c+d)k^3t$. We shall further denote $S\equiv a+b+c+d$. Now we can equate to 0 the coefficients for each $e^{in\theta}$ starting from eq. (\ref{sNLS:bona}):
\begin{equation}
    \begin{split}
        &\left(\frac{\partial}{\partial t}-ink+\frac{S}{2}ink^3\right)\eta_n+\left(\frac{\partial}{\partial x}+ink\right)u_n+\left(\frac{\partial}{\partial x}+ink\right)\sum\limits_{j=-\infty}^{\infty}u_j\eta_{n-j}\\
        &+a\left(\frac{\partial}{\partial x}+ink\right)^3u_n-b\left(\frac{\partial}{\partial x}+ink\right)^2\left(\frac{\partial}{\partial t}-ink+\frac{S}{2}ink^3\right)\eta_n=0\\
        \\
      &\left(\frac{\partial}{\partial t}-ink+\frac{S}{2}ink^3\right)u_n+\left(\frac{\partial}{\partial x}+ink\right)\eta_n+\sum\limits_{j=-\infty}^{\infty}u_j\left(\frac{\partial}{\partial x}+ink\right)u_{n-j}\\
      &+c\left(\frac{\partial}{\partial x}+ink\right)^3\eta_n-d\left(\frac{\partial}{\partial x}+ink\right)^2\left(\frac{\partial}{\partial t}-ink+\frac{S}{2}ink^3\right)u_n=0
    \end{split}
    \label{Bona_expand}
\end{equation}
Let us introduce a small parameter $\varepsilon$. We will now seek for the modulation solutions of the form:
\begin{equation}
    \begin{split}
        &u_n=\varepsilon^{\alpha_n}v_n(\xi,\tau)\\
        &\eta_n=\varepsilon^{\alpha_n}\varphi_n(\xi,\tau)
    \end{split}
\end{equation}
with:
\begin{equation}
    \begin{split}
        &\xi=\varepsilon\left(x-t+\frac{3S}{2}k^2t\right)\\
        &\tau=\varepsilon^2t
    \end{split}
\end{equation}
We choose $\alpha_0=2$ and $\alpha_n=\alpha_{-n}=n$ for $n\ge 1$. The differential operators rewrite:
\begin{equation}
    \partial_x\to \varepsilon\partial_{\xi}
\end{equation}
\begin{equation}
    \partial_t\to \varepsilon^2\partial_{\tau}+\varepsilon\left(\frac{3S}{2}-1\right)\partial_{\xi}
\end{equation}
When taking $n=1$ and order three in $\varepsilon$ we obtain from system (\ref{Bona_expand}):
\begin{equation}
    \begin{split}
        &\partial_{\tau}\varphi_1+ik(v_1\varphi_0+v_0\varphi_1+v_2\varphi_{-1}+v_{-1}\varphi_2)+3aik\partial_{\xi\xi} v_1\\
        &-b\left(\left(\frac{iSk^3}{2}-ik\right)\partial_{\xi \xi}\varphi_1 + 2ik\left(\frac{3S}{2}-1\right)\partial_{\xi\xi}\varphi_1 -k^2\partial_{\tau}\varphi_1 \right)=0\\
        \\
        &\partial_{\tau}v_1+2ik(v_1v_0+v_2v_{-1})+3cik\partial_{\xi \xi}\varphi_1\\
        &-d\left(\left(\frac{iSk^3}{2}-ik\right)\partial_{\xi \xi}v_1 + 2ik\left(\frac{3S}{2}-1\right)\partial_{\xi\xi}v_1 -k^2\partial_{\tau}v_1 \right)=0
    \end{split}
    \label{sNLS:seq}
\end{equation}
For $n=0$ and order three in $\varepsilon$ we obtain:
\begin{equation}
    \begin{split}
        & \left(\frac{3S}{2}-1\right)\partial_{\xi}\varphi_0+\partial_{\xi}v_0+\partial_{\xi}(v_1 \varphi_{-1}+v_{-1}\varphi_1) =0\\
        &\left(\frac{3S}{2}-1\right)\partial_{\xi}v_0+\partial_{\xi}\varphi_0+\partial_{\xi} (v_1v_{-1})=0
    \end{split}
\end{equation}
By integrating this system we arrive at (setting all constants equal to 0 in the RHS):
\begin{equation}
    \begin{split}
        & \left(\frac{3S}{2}-1\right)\varphi_0+v_0 +v_1 \varphi_{-1}+v_{-1}\varphi_1=0\\
        &\left(\frac{3S}{2}-1\right)v_0+\varphi_0 + v_1v_{-1}=0
    \end{split}
\end{equation}
The solution of this system is:
\begin{equation}
    \begin{split}
        &v_0=\frac{ (4-6 S) v_{-1}v_1+4 (v_{-1}\varphi _1+ v_1 \varphi
   _{-1})}{3 S (3 S-4)}\\
        &\varphi_0=\frac{4 v_{-1} v_1+(4-6S) \left(v_1 \varphi
   _{-1}+v_{-1} \varphi _1\right)}{3S (3S-4)}
    \end{split}
\end{equation}
For $n=2$ and order $\varepsilon^2$ we have:
\begin{equation}
    \begin{split}
        &(-2ik+Sik^3)\varphi_2+2ik v_2+2ik v_1\varphi_1
        +a(2ik)^3v_2-b(2ik)^2(-2ik+Sik^3)\varphi_2=0\\
        &(-2ik+Sik^3)v_2+2ik \varphi_2+2ik v_1^2
        +c(2ik)^3\varphi_2-d(2ik)^2(-2ik+Sik^3)v_2=0
    \end{split}
\end{equation}
The solutions of this system are:
\begin{equation}
    \begin{split}
        &\varphi_2=\frac{v_1 \left(v_1 \left(4-16 a k^2\right)-2 \varphi _1 \left(4 d k^2+1\right) \left(k^2 S-2\right)\right)}{k^2 \left(a
   \left(16-64 c k^2\right)+4 b \left(4 d k^2+1\right) \left(k^2 S-2\right)^2+16 (c+d)+k^2 S \left(4 d \left(k^2
   S-4\right)+S\right)-4 S\right)}\\
   &v_2=\frac{4 v_1 \varphi _1 \left(1-4 c k^2\right)-2 v_1^2 \left(4 b k^2+1\right) \left(k^2 S-2\right)}{k^2 \left(a \left(16-64 c
   k^2\right)+4 b \left(4 d k^2+1\right) \left(k^2 S-2\right)^2+16 (c+d)+k^2 S \left(4 d \left(k^2 S-4\right)+S\right)-4
   S\right)}
    \end{split}
\end{equation}

Let us show two particular cases:

\begin{enumerate}
    \item $a=1/3, b=c=d=0$\\
    In this case the general system is reduced to the completely integrable Kaup-Boussinesq system.
    Equations (\ref{sNLS:seq}) writes:
\begin{equation}
    \begin{split}
        &\partial_{\tau}\varphi_1+ik(v_1\varphi_0+v_0\varphi_1+v_2\varphi_{1}^*+v_{1}^*\varphi_2)+ik\partial_{\xi\xi} v_1=0\\
        &\partial_{\tau}v_1+2ik(v_1v_0+v_2v_{1}^*)=0
    \end{split}
\end{equation}
We have:
\begin{equation}
\begin{split}
    &\varphi_0=-\frac{2}{3} \left(\varphi _{1}^* v_1+v_{1}^* \varphi _1+2 v_{1}^* v_1\right)\\
    &v_0=-\frac{2}{3} \left(2\varphi _{1}^* v_1+2v_{1}^* \varphi _1+v_{1}^* v_1\right)\\
    &\varphi_2=\frac{v_1 \left(\left(4-\frac{16 k^2}{3}\right) v_1-2 \left(\frac{k^2}{3}-2\right) \varphi _1\right)}{k^2 \left(\frac{16}{3}+\frac{k^2}{9}-\frac{4}{3}\right)}\\
   &v_2=\frac{4 v_1 \varphi _1 -2 \left(\frac{k^2}{3}-2\right) v_1^2}{k^2 \left(\frac{16}{3}+\frac{k^2}{9}-\frac{4}{3}\right)}
\end{split}
\end{equation}
By replacing we get:
\begin{equation}
    \begin{split}
        &\partial_{\tau}\varphi_1+ik(A_{1}v_1^2\varphi_1^*+A_{2}|v_1|^2\varphi_1+A_{3}|v_1|^2v_1+A_{4}|\varphi_1|^2v_1+A_{5}\varphi_1^2v_1^*)+ik\partial_{\xi\xi} v_1=0\\
        &\partial_{\tau}v_1+2ik(B_{1}v_1^2\varphi_1^*+B_{2}|v_1|^2\varphi_1+B_{3}|v_1|^2v_1)=0
    \end{split}
\end{equation}
where:
$$A_1=-\frac{2k^4+90 k^2-108}{3 k^2 (k^2+36)}, A_2=-\frac{4k^4+162 k^2-108}{3k^2 (k^2+36)}, A_3=-\frac{4 (-27 + 72 k^2 + k^4)}{3 k^2 (k^2+36)},$$
$$A_4=-\frac{4k^4+144k^2-108}{3 k^2 (k^2+36)}, A_5=-\frac{4}{3}, B_1=-\frac{4}{3}, B_2=-\frac{4k^4+144k^2-108}{3 k^2 (k^2+36)}, B_3=-\frac{2(-54 + 45 k^2 + k^4)}{3 k^2 (36 + k^2)}$$
{\bf Remark:} 

We expect that this {\it dispersionless} system to be also completely integrable.

\item $a=c=1/3, b=d=0$\\ This is a non-integrable coupled KdV system. The resulting envelope system (\ref{sNLS:seq}) writes:
    \begin{equation}
        \begin{split}
        &\partial_{\tau}\varphi_1+ik(v_1\varphi_0+v_0\varphi_1+v_2\varphi_{1}^*+v_{1}^*\varphi_2)+ik\partial_{\xi\xi} v_1=0\\
        &\partial_{\tau}v_1+2ik(v_1v_0+v_2v_{1}^*)+ik\partial_{\xi \xi}\varphi_1=0
        \end{split}
    \end{equation}
    Where:
    \begin{equation}
        \begin{split}
            &\varphi_0=-v_{1}^* v_1\\
            &v_0=-v_1 \varphi _{1}^*-v_{1}^* \varphi _1\\
            &\varphi_2=\frac{v_1 \left(\left(4-\frac{16 k^2}{3}\right) v_1-2 \left(\frac{2 k^2}{3}-2\right) \varphi _1\right)}{k^2 \left(\frac{4
   k^2}{9}+\frac{1}{3} \left(16-\frac{64 k^2}{3}\right)+\frac{8}{3}\right)}\\
   &v_2=\frac{4 \left(1-\frac{4 k^2}{3}\right) v_1 \varphi _1-2 \left(\frac{2 k^2}{3}-2\right) v_1^2}{k^2 \left(\frac{4
   k^2}{9}+\frac{1}{3} \left(16-\frac{64 k^2}{3}\right)+\frac{8}{3}\right)}
        \end{split}
    \end{equation}
By replacing we get:
\begin{equation}
    \begin{split}
        &\partial_{\tau}\varphi_1+ik(A_{1}|v_1|^2v_1+A_{2}|\varphi_1|^2v_1+A_{3}\varphi_1^2v_1^*+A_{4}v_1^2\varphi_1^*+A_{5}|v_1|^2\varphi_1)+ik\partial_{\xi\xi} v_1=0\\
        &\partial_{\tau}v_1+2ik(B_{1}v_1^2\varphi_1^*+B_{2}|v_1|^2\varphi_1+B_{3}|v_1|^2v_1)+ik\partial_{\xi \xi}\varphi_1=0
    \end{split}
\end{equation}
where
$$A_3=B_1=-1, A_1=A_2=B_2=-\frac{(3 - 10 k^2 + 5 k^4)}{k^2(5k^2-6)},  A_4=A_5=B_3=\frac{k^2-3}{k^2(5k^2-6)}$$

\end{enumerate}

\section{Solitary wave solution}
In order to compute solitary wave solution for the general system (\ref{sNLS:seq}) we have to write it in a more condensed form.
Let us write again the relation defining $v_0, v_2, \varphi_0, \varphi_2$:

$$v_0=\alpha_1|v_1|^2+\alpha_2 (v_1\varphi_1^*+v_1^*\varphi_1)$$
$$\varphi_0=\beta_2|v_1|^2+\beta_1 (v_1\varphi_1^*+v_1^*\varphi_1)$$
(in which $\alpha_1=\beta_1=(4-6S)/(9S^2-12S), \alpha_2=\beta_2=4/(9S^2-12S)$)
$$\varphi_2=\gamma_1v_1^2+\gamma_2\varphi_1v_1$$
$$v_2=\delta_2v_1\varphi_1+\delta_2v_1^2$$ with coefficients given by (16).
Accordingly the nonlinear terms of the coupled NLS will be:
$$v_1\varphi_0+v_0\varphi_1+v_2\varphi_1^*+v_1^*\varphi_2=(\beta_1+\delta_2)\varphi_1^*v_1^2+(\beta_1+\alpha_1+\gamma_2)\varphi_1|v_1|^2+(\beta_2+\gamma_1)v_1|v_1|^2+(\alpha_2+\delta_1)v_1|\varphi_1|^2+\alpha_2v_1^*\varphi_1^2$$
and
$$v_1v_0+v_2v_1^*=(\alpha_1+\delta_2)v_1|v_1|^2+(\alpha_2+\delta_1)\varphi_1|v_1|^2+\alpha_2 v_1^2\varphi_1^*$$

Our system (\ref{sNLS:seq}) will have the form:
$$
 i\dot\varphi_1-\frac{3ak}{1+bk^2}v_1''+\frac{b\sigma_0}{1+bk^2}\varphi_1''-\frac{k}{2(1+bk^2)}((\beta_1+\delta_2)\varphi_1^*v_1^2+(\beta_1+\alpha_1+\gamma_2)\varphi_1|v_1|^2+(\beta_2+\gamma_1)v_1|v_1|^2+
 $$
\begin{equation}
 +(\alpha_2+\delta_1)v_1|\varphi_1|^2+\alpha_2v_1^*\varphi_1^2)=0
\end{equation}
\begin{equation}
 i\dot v_1-\frac{3ck}{1+dk^2}\varphi_1''+\frac{d\sigma_0}{1+dk^2} v_1''-\frac{k}{(1+bk^2)}((\alpha_1+\delta_2)v_1|v_1|^2+(\alpha_2+\delta_1)\varphi_1|v_1|^2+\alpha_2 v_1^2\varphi_1^*)=0
\end{equation}
where $\sigma_0=\frac{1}{2}Sk^3-3Sk-3k$. The system can be written generally (with arbitrary coefficients)

\begin{equation}\label{nls1}
i\dot\varphi_1-s_1v_1''+s_2\varphi_1''-s_3\varphi_1^*v_1^2-s_4\varphi_1|v_1|^2-s_5v_1|v_1|^2-s_6v_1|\varphi_1|^2-s_7v_1^*\varphi_1^2=0
\end{equation}
\begin{equation}\label{nls2}
 i\dot v_1-\sigma_1\varphi_1''+\sigma_2 v_1''-\sigma_3v_1|v_1|^2-\sigma_4\varphi_1|v_1|^2-\sigma_5 v_1^2\varphi_1^*=0
\end{equation}

In order to find the solitary wave solution for this system we will implement Hirota bilinear form. However the system is not integrable and thus we expect to have a {\it multilinear} (not a bilinear) form. 
Lets consider the following susbtitutions:
$$\varphi_1=G/F,\quad v_1=H/F,\quad G, H, {\rm complex}, F, {\rm real}$$ 

and the following properties which are necessary for our problem
$$\partial_x(G/F)=\frac{D_xG\cdot F}{F^2},\quad \partial_x^2(G/F)=\frac{D_x^2G\cdot F}{F^2}-\frac{G}{F}\frac{D_x^2F\cdot F}{F^2}$$

Introducing in the system (\ref{nls1}), (\ref{nls2}) we find
$$F(iD_{\tau}G\cdot F-s_1D_{\xi}^2H\cdot F+s_2D_{\xi}^2G\cdot F)+H(s_1D_{\xi}^2F\cdot F-s_3G^*H-s_4H^*G)+G(-s_2D_{\xi}^2F\cdot F-s_4|H|^2-s_7H^*G)=0$$

$$F(iD_{\tau}H\cdot F-\sigma_1D_{\xi}^2G\cdot F+\sigma_2D_{\xi}^2H\cdot F)+G(\sigma_1D_{\xi}^2F\cdot F-\sigma_5|H|^2)+H(-\sigma_2D_{\xi}^2F\cdot F-\sigma_3|H|^2-\sigma_4HG^*)=0$$

We try now the ansatz used for ordinary nonlinear Schrödinger equation, namely
$$G=\alpha e^{\eta},\quad H=\beta e^{\eta}, \quad F=1+b e^{\eta+\eta^*}$$
where $\eta=K\xi+\Omega \tau$ is a {\it complex} phase with {\it both} $K$ and $\Omega$ complex numbers. Introducing in the above trilinear forms we will obtain an algebraic system for dispersion relation of the nonlinear wave  $\Omega=\Omega(K)$ and the amplitudes $\alpha,\beta, b$. 
For simplification we isolate the first bilinear parantheses which mimics the linearised part of the system and try to find the dispersion:
$$iD_{\tau}G\cdot F-s_1D_{\xi}^2H\cdot F+s_2D_{\xi}^2G\cdot F=0$$
$$iD_{\tau}H\cdot F-\sigma_1D_{\xi}^2G\cdot F+\sigma_2D_{\xi}^2H\cdot F=0$$

First bilinear equation gives:
$$e^{\eta}(i\alpha\Omega-s_1\beta K^2+s_2\alpha K^2)+e^{2\eta+\eta^*}(-i\alpha b\Omega^*-s_1 b \beta K^{*2}+s_2b\alpha K^{*2})=0$$
$$e^{\eta}(i\beta\Omega-\sigma_1\alpha K^2+\sigma_2\beta K^2)+e^{2\eta+\eta^*}(-i\beta b\Omega^*-\sigma_1\alpha b K^{*2}+\sigma_2 b \beta K^{*2})=0$$
One can see immediately that in every equation the second term is the complex conjugate of the first and multiplied by $b$ (provided $\alpha$, $\beta$ are real when $s_1$ and $\sigma_1$ are not zero). So we can write the compatbility for the system in $\alpha, \beta$ and obtain the following
$$
{\rm det}
\left(
\begin{array}{cc}
i\Omega+s_2K^2&-s_1K^2\\
-\sigma_1K^2&i\Omega+\sigma_2K^2
\end{array}\right)=0
$$
and the dispersion (here $\Omega=\Omega_R+i\Omega_I, K=K_R+iK_I$)
$$\Omega(K)=\frac{iK^2}{2}\left(\sigma_2+s_2\pm \sqrt{(s_2+\sigma_2)^2+4(s_1\sigma_1-s_2\sigma_2)}\right)$$

Next we have to cancel the trilinear leftover terms:
$$H(s_1D_x^2F\cdot F-s_3G^*H-s_4H^*G)+G(-s_2D_x^2F\cdot F-s_4|H|^2-s_7H^*G)=0$$
$$G(\sigma_1D_x^2F\cdot F-\sigma_5|H|^2)+H(-\sigma_2D_x^2F\cdot F-\sigma_3|H|^2-\sigma_4HG^*)=0$$
Introducing the ansatz we obtain the following algebraic system;
$$2 s_1 b \beta (K+K^*)^2 - \beta^2\alpha (s_3+s_4) - 2 \alpha s_2 b (K+K^*)^2 - \alpha \beta^2 s_4 - s_7\beta \alpha^2=0$$
$$
2 \sigma_1 b \alpha (K+K^*)^2 - \beta^2 \alpha \sigma_5 - 
 2 \beta \sigma_2 b (K+K^*)^2 - \beta^3\sigma_3 - \beta^2
\alpha \sigma_4=0
$$
The unknowns are $\alpha, \beta, b$ where $b$ must be real (since $F$ is real). Since we have only two equations, $\alpha$ can be arbitrary (we can put $\alpha=1$)  and solve for $b, \beta$. With the help of MATHEMATICA one can find the expressions for $b, \beta$ which unfortunately are very long. However for the case of $s_1=\sigma_1=0$ (which means $a=c=0$ in the initial system) we find the following
$$\beta=\alpha\frac{s_2(\sigma_4+\sigma_5)-s_7\sigma_2}{s_3\sigma_2+2s_4\sigma_3-s_2\sigma_3}$$
$$b=-\frac{\alpha^2}{2(K+K^*)^2}\frac{s_2(\sigma_4+\sigma_5)-s_7\sigma_2}{(s_3\sigma_2+2s_4\sigma_3-s_2\sigma_3)^2}$$

So finally the solitary wave (here $\alpha$ is arbitary and {\it can} be real or purely imaginary)
$$\varphi_1=\alpha e^{i(K_I\xi+\Omega_I\tau)}\frac{e^{K_R\xi+\Omega_R\tau}}{1+be^{2K_R\xi+2\Omega_R\tau}},\quad v_1=\beta e^{i(K_I\xi+\Omega_I\tau)}\frac{e^{K_R\xi+\Omega_R\tau}}{1+be^{2K_R\xi+2\Omega_R\tau}}$$

\end{document}